\documentclass[journal]{IEEEtran}
 
\usepackage{graphicx}
\usepackage{booktabs}
\usepackage{subfigure}
\usepackage{overpic}
\usepackage{amsmath}
\usepackage{flushend}
\usepackage{amssymb}
\usepackage{multirow}
\usepackage{makecell}
\usepackage{color, soul}
\soulregister\cite7

\usepackage[ruled]{algorithm2e}
\usepackage{hyperref}
\hypersetup{
	colorlinks=true,
	linkcolor=blue,
	urlcolor=blue,
	citecolor=blue}

\usepackage{colortbl}
\usepackage[dvipsnames]{xcolor}
\definecolor{Bg}{HTML}{e0f1ff}
\definecolor{Dh}{HTML}{ffb3a7}

\begin{document}

\title{Global and Local Attention-Based Transformer for Hyperspectral Image Change Detection}
\author{Ziyi Wang, Feng Gao, \textit{Member, IEEE}, Junyu Dong, \textit{Member, IEEE}, and Qian Du, \textit{Fellow, IEEE}
\thanks{This work was supported in part by the National Science and Technology Major Project under Grant 2022ZD0117201 and in part by the Natural Science Foundation of Qingdao under Grant 23-2-1-222-ZYYD-JCH. \textit{(Corresponding author: Feng Gao)}

Ziyi Wang, Feng Gao, Junyu Dong are with the School of Information Science and Engineering, Ocean University of China, Qingdao 266100, China. 

Qian Du is with the Department of Electrical and Computer Engineering, Mississippi State University, Starkville, MS 39762 USA.}}

\markboth{IEEE Geoscience and Remote Sensing Letters}%
{Shell }

\maketitle

\begin{abstract}
Recently Transformer-based hyperspectral image (HSI) change detection methods have shown remarkable performance. Nevertheless, existing attention mechanisms in Transformers have limitations in local feature representation. To address this issue, we propose \textbf{G}lobal and \textbf{L}ocal \textbf{A}ttention-based Trans\textbf{former} (GLAFormer), which incorporates a global and local attention module (GLAM) to combine high-frequency and low-frequency signals. Furthermore, we introduce a cross-gating mechanism, called cross-gated feed-forward network (CGFN), to emphasize salient features and suppress noise interference. Specifically, the GLAM splits attention heads into global and local attention components to capture comprehensive spatial-spectral features. The global attention component employs global attention on downsampled feature maps to capture low-frequency information, while the local attention component focuses on high-frequency details using non-overlapping window-based local attention. The CGFN enhances the feature representation via convolutions and cross-gating mechanism in parallel paths. The proposed GLAFormer is evaluated on three HSI datasets. The results demonstrate its superiority over state-of-the-art HSI change detection methods. The source code of GLAFormer is available at \url{https://github.com/summitgao/GLAFormer}.

\end{abstract}


\IEEEpeerreviewmaketitle

\section{Introduction}

\IEEEPARstart{H}{yperspectral} image (HSI) change detection stands as a crucial task within the field of remote sensing, focusing on the identification of altered areas by comparing hyperspectral images obtained at different times. The exceptional spectral resolution of HSIs facilitates accurate detection of changes in ground objects \cite{ding22grsl}. As such, HSI change detection has been widely applied in various domains, including damage assessment \cite{entcheva04ijrs}, land cover analysis \cite{8738052} and urban expansion monitoring\cite{9773336}.

Traditional methods in HSI change detection primarily employed techniques, such as change vector analysis \cite{cva06tgrs} and Tucker decomposition \cite{hou21jstars}, to analyze the spectral changes in multi-temporal images. Canty et al. \cite{niemeyer2003pixel} were the first to introduce the Multivariate Alteration Detection (MAD) method, which is based on Canonical Correlation Analysis and is designed for unsupervised change detection in vegetation using multi-temporal hyperspectral images. Later, Nielsen \cite{nielsen2007regularized} enhanced this approach by developing the Iteratively Reweighted MAD (IR-MAD) algorithm. However, these methods encountered limitations in threshold selection and reduced robustness in complex scenarios \cite{hasanlou2018hyperspectral}. Recently, convolutional neural networks (CNNs) \cite{lecun1998gradient} have been proven especially effective in extracting representative features from multi-temporal HSIs. Saha et al. \cite{deepcva} proposed a method called deep CVA by combining CNNs with change vector analysis (CVA).

More recently, Transformers, which rely entirely on self-attention, have gained popularity in computer vision tasks, such as image classification \cite{vit} and object detection \cite{dert}. Transformers have the advantage of capturing global dependencies and exhibit better performance in handling long-range dependencies compared to CNNs. This potential has prompted researchers to explore the attention mechanisms for HSI change detection. Song et al. \cite{song2022csanet} enhanced the feature representation of multi-temporal HSIs by introducing cross-temporal interaction symmetric attention. Furthermore, Ding et al. \cite{ding22grsl} introduced the Transformer encoder for HSI change detection, leveraging self-attention with a global receptive field to enhance the recognition of changes. In \cite{wang2022vtc}, a Transformer-based multi-scale feature fusion model was proposed for change detection.

\begin{figure*}[ht]
\centering
\includegraphics [width=6.8in]{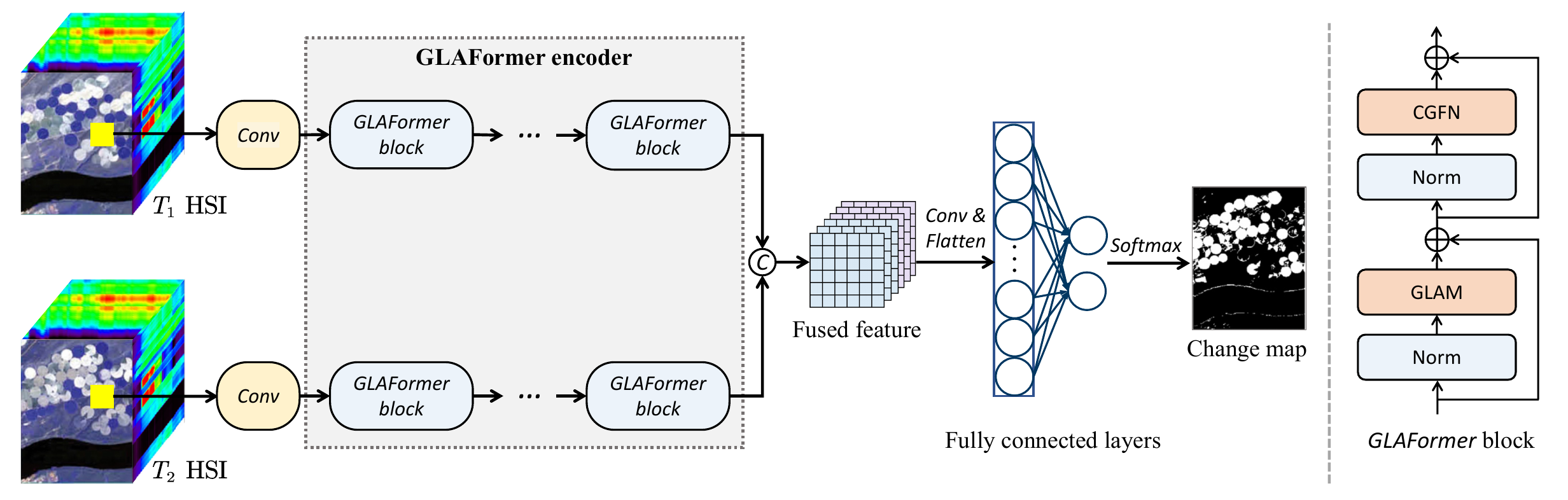}
\caption{Overview of the proposed GLAFormer. In the GLAFormer block, high-frequency and low-frequency signals are fused through global and local attention module (GLAM). The ability to capture local context is enhanced through the cross-gated feed-forward network (CGFN). The final output is the Change Map, where white pixels represent the changed areas.} 
\label{img_frame}
\end{figure*}

Although existing Transformer-based methods for HSI change detection have achieved promising performance, they still suffer from two limitations: 1) \textbf{Insufficient local-level representations modeling.} Transformers tend to pay more attention to global features. However, global and local features serve distinct roles in encoding HSI patterns. The emphasis on global features in Transformers leads to the loss of certain local features, which results in a degradation of change detection performance. 2) \textbf{Limited non-linear feature transformation.} Feed-Forward Network (FFN) is commonly used to process the output from the attention layer in Transformer, enabling non-linear feature transformation for the input of the subsequent attention layer. However, existing methods are limited in non-linear feature representation and are susceptible to noise interference.

To overcome the above limitations, we propose a \textbf{G}lobal and \textbf{L}ocal \textbf{A}ttention-based Trans\textbf{former} for HSI change detection, GLAFormer for short. Specifically, to enhance the local-level feature representations, we have designed a global and local attention module (GLAM) to encode both low-frequency and high-frequency signals. Furthermore, to augment the non-linear feature transformation, we propose the cross-gated feed-forward network (CGFN) to amplify the salient information while suppressing noise. Extensive experiments on three HSI change detection datasets demonstrate the superiority of our proposed GLAFormer.

Our main contributions can be summarized as follows:

\begin{itemize}

\item We propose the GLAM as an enhancement to the self-attention mechanism. This module combines high-frequency and low-frequency signals to achieve a more comprehensive spatial-spectral feature representation for change detection.

\item We develop the CGFN to improve the non-linear feature transformation within Transformers. This network amplifies important information and mitigates noise interference.

\item Extensive experimental results demonstrate that the proposed GLAFormer outperforms state-of-the-art methods. The codes will be released to the remote sensing community.

\end{itemize}

\section{Methodology}

The overall architecture of our proposed GLAFormer is shown in Fig. \ref{img_frame}. Two hyperspectral images ($T_1$ and $T_2$) captured at different times are passed to the GLAFormer. Firstly, two patches from the multi-temporal HSIs of the same geographical area are extracted. Then, both patches are fed into two parallel GLAFormer encoders to extract informative and robust features. Next, the learned features from the two paths are fused. Finally, the fused features are transformed by several convolutional and fully connected layers for change detection.  

As shown in the right part of Fig. \ref{img_frame}, the GLAFormer block consists of two key modules: GLAM and CGFN, which are detailed as below.

\begin{figure*}[ht]
\centering
\includegraphics [width=6.5in]{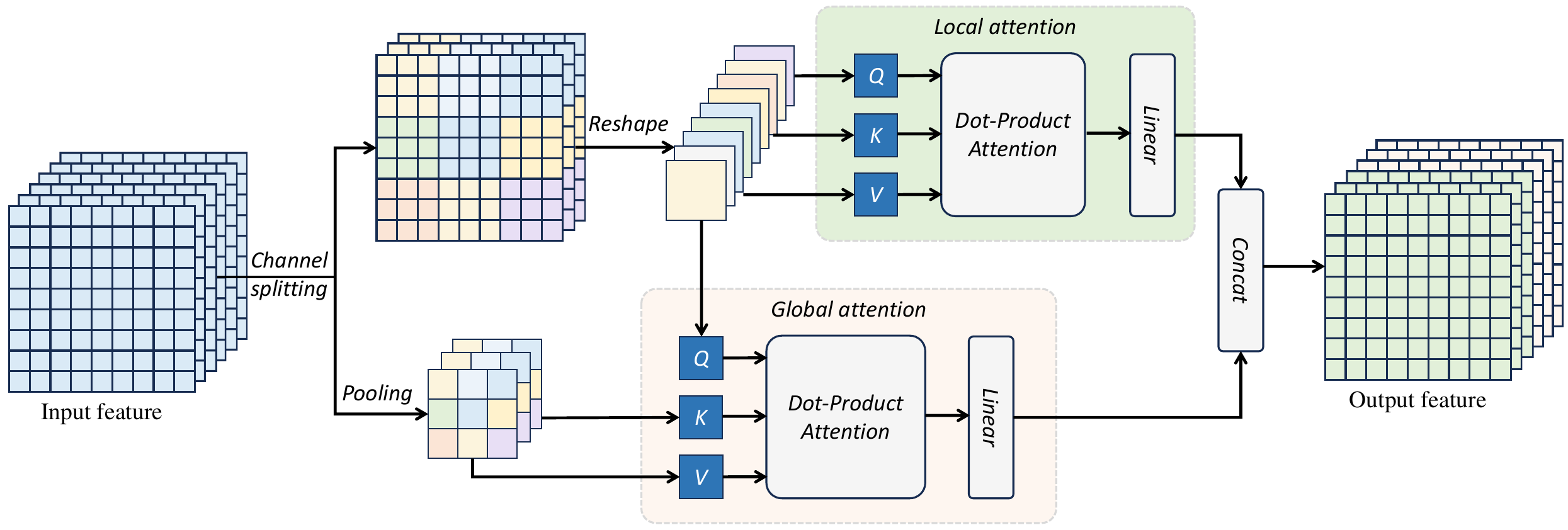}
\caption{Illustration of global and local attention module (GLAM). The module consists of two branches: global attention and local attention. The global attention branch captures the long-range dependencies of the input, while the local attention branch computes the detailed local feature dependency. Then, the global and local features are fused by concatenation.}
\label{img_HiLo}
\end{figure*}

\subsection{Global and Local Attention Module (GLAM)}

As depicted in Fig. \ref{img_HiLo}, the GLAM consists of two branches: global attention and local attention. The global attention captures the global dependencies of the input, while the local attention branch computes the detailed local feature dependency. The global and local features are fused by concatenation.

\textbf{Local attention.} The local attention branch encodes high-frequency features via local window self-attention, which applies self-attention mechanism to local windows of feature maps. As shown in Fig  \ref{img_HiLo}, a local window refers to a $3\times3$ region in the feature maps. These local windows are evenly partitioned in a non-overlapping manner. 

The feature within each window is of size $s \times s \times d$. Here, $s$ is the size of the window, and $d$ is the feature dimension. The feature within each window is reshaped into $\mathbf{X}_{\text{l}} \in \mathbb{R}^{d \times s^2}$. Next, a $1\times1$ convolution is applied to enhance the input and obtain the query $\mathbf{Q}_{\text{l}}$, key $\mathbf{K}_{\text{l}}$, and value $\mathbf{V}_{\text{l}}$. The local window attention is defined as:

\begin{equation}
\hat{\mathbf{X}}_{\text{l}} = \mathbf{V}_{\text{l}} \cdot \text{Softmax}\left(\frac{\mathbf{K}_{\text{l}}^T \mathbf{Q}_{\text{l}}}{\sqrt{D}}\right),
\end{equation}
where $\hat{\mathbf{X}}_{\text{l}}$ is the output feature from the local attention branch. $\mathbf{Q}_{\text{l}}$, $\mathbf{K}_{\text{l}}$, and $\mathbf{V}_{\text{l}}$ are the tensors after the $1\times1$ convolutions specific to the local attention branch, and $D$ is the number of hidden dimensions for a single head in the local attention.

\textbf{Global attention.} The global attention branch captures low-frequency features by applying the attention mechanism over pooled feature maps. As illustrated in Fig. \ref{img_HiLo}, the input feature is evenly partitioned into $3\times3$ windows in a non-overlapping manner. To effectively capture global information, average pooling is employed on each window to obtain the average-pooled feature map $\mathbf{Z} \in \mathbb{R}^{d \times 9}$. Next, $\mathbf{Z}$ is transformed to key $\mathbf{K}_{\text{g}} \in \mathbb{R}^{d \times 9}$ and value $\mathbf{V}_{\text{g}} \in \mathbb{R}^{d \times 9}$. To ensure complete and unchanged information access, the global attention uses queries $\mathbf{Q}_{\text{l}} \in \mathbb{R}^{d \times 81}$ from the original feature map. This approach is consistent with that of local attention. To generate the output features $\hat{\mathbf{Z}}_{\text{g}}$, the standard self-attention is applied on $\mathbf{Q}_{\text{l}}$, $\mathbf{K}_{\text{g}}$, and $\mathbf{V}_{\text{g}}$:

\begin{equation}
\hat{\mathbf{Z}}_{\text{g}} = \mathbf{V}_{\text{g}} \cdot \text{Softmax}\left(\frac{\mathbf{K}_{\text{g}}^T \mathbf{Q}_{\text{l}}}{\sqrt{D}}\right).
\end{equation}

\textbf{Channel splitting and merging.} The input features are evenly split along the channel dimension before entering the global and local attention branches, reducing complexity and boosting GPU throughput.  This splitting also decomposes the learnable parameters into smaller matrices, reducing the model's parameter count. These two sets of features are separately fed into the local and global attention branches, respectively. To produce the output of GLAM, the output features $\hat{\mathbf{X}}_{\text{l}}$ from the local attention branch and the output features $\hat{\mathbf{Z}}_{\text{g}}$ from the global attention branch are concatenated as:

\begin{equation}
\mathbf{F}_{\text{GLAM}} = \text{concat}[\hat{\mathbf{X}}_{\text{l}}; \hat{\mathbf{Z}}_{\text{g}}].
\end{equation}

\subsection{Cross-Gated Feed-Forward Network (CGFN)}

\begin{figure}
  \centering
  \includegraphics[width=3in]{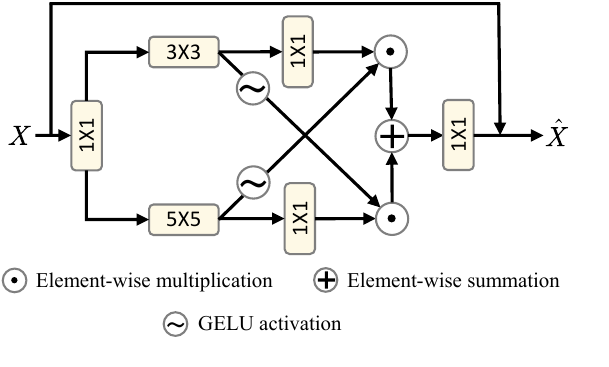}
  \caption{The architecture of cross-gated feed-forward network (CGFN).}
  \label{img_dgfn}
\end{figure}

To enhance the non-linear feature transformation in Transformers, the CGFN is proposed, which incorporates the gating mechanism and multi-scale convolution into the existing feed-forward network.

As shown in Fig. \ref{img_dgfn}, the proposed CGFN consists of two parallel paths. In each path, depth-wise convolutions with different sizes of kernels are employed to enhance the multi-scale feature extraction. Then, the gating mechanism is used to filter the less informative features in each path. The useful features passing through the gates are fused with the original features from another path. The fused features from the two paths are combined via element-wise summation. Given input $\mathbf{Y}$, the CGFN can be defined as:
\begin{equation}
\begin{split}
\mathrm{Gating}(\mathbf{Y}) = \phi\left(W_{3\times3} W_{1\times1}^0 \mathbf{Y}\right) \odot\left(W_{1\times1}^{2}W_{5\times5} W_{1 \times 1}^0 \mathbf{Y}\right) ~~~\\
+\phi\left(W_{5\times5} W_{1\times1}^0 \mathbf{Y} \right) \odot \left(W_{1\times1}^{3} W_{3 \times 3} W_{1 \times 1}^0 \mathbf{Y}\right), ~~
\end{split}
\end{equation}
\begin{equation}
\hat{\mathbf{Y}} =W_{1\times1}^4 \mathrm{Gating}(\mathbf{Y})+\mathbf{Y},
\end{equation}
where $\hat{\mathbf{Y}}$ denotes the output features, $\mathrm{Gating}(\cdot)$ denotes the cross-gated mechanism,  $\phi$ is the GELU activation function and $\odot$ represents the element-wise multiplication operation. The gating mechanism and the multi-scale convolutions  amplify important information and mitigate noise interference. 

\begin{figure}
  \centering
  \includegraphics[width=3in]{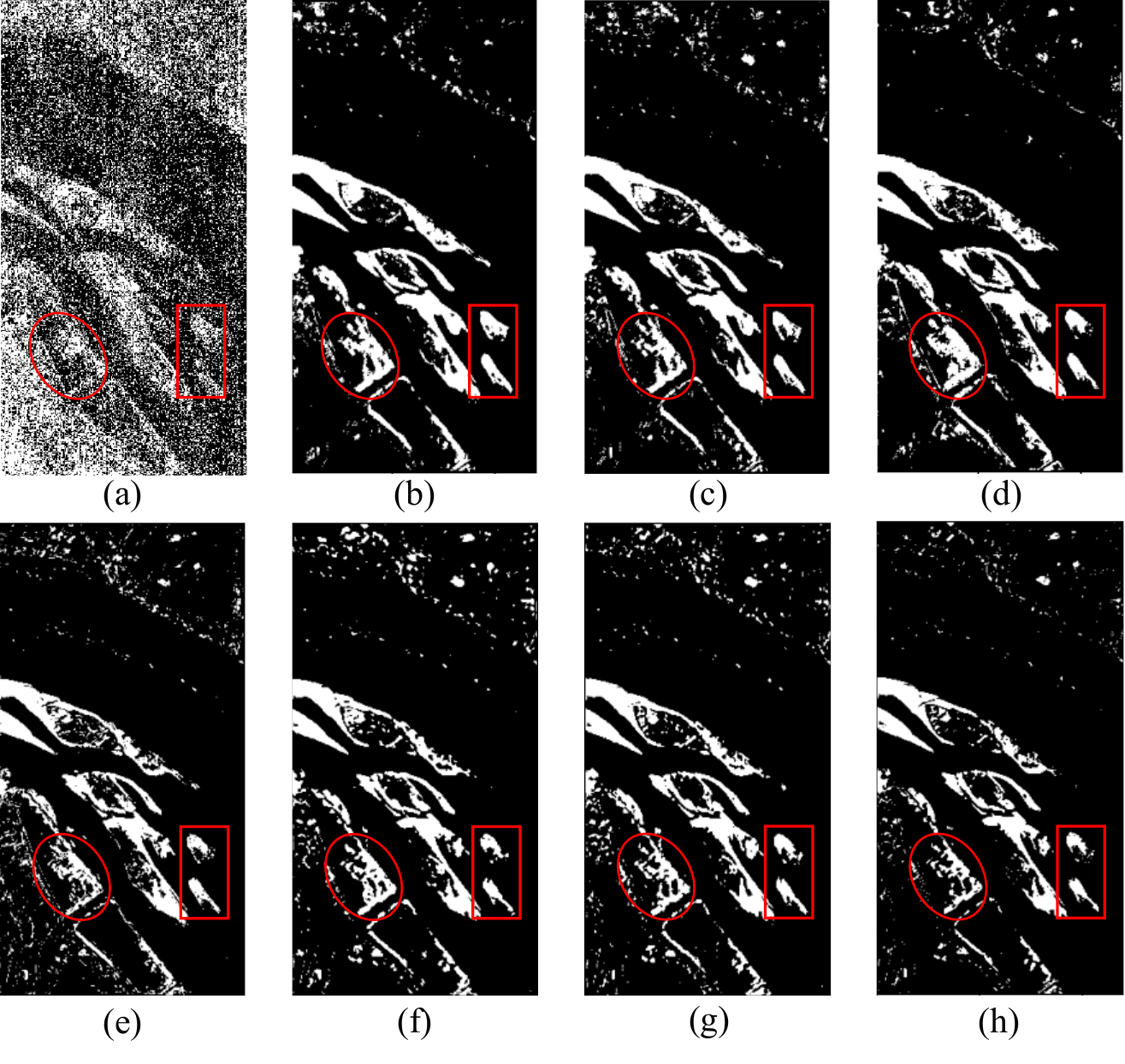}
  \caption{Change detection results of different methods on River dataset. (a) IR-MAD. (b) SSA-SiamNet. (c) SSCNN-S. (d) CDFomer. (e) SSTFormer. (f) CSDBF. (g) Proposed GLAFormer. (h) Ground truth.}
  \label{fig_river}
\end{figure}

\section{Experimental Results and Analysis}

\subsection{Dataset and Experimental Setting}

We evaluate the performance of our GLAFormer through extensive experiments on three widely recognized multi-temporal hyperspectral datasets. These datasets were sourced from the Hyperion sensor onboard the EO-1 satellite. Specifically, the first dataset, referred to as the River dataset \cite{river}, comprises imagery of a river in Jiangsu Province, China. The second dataset is the Farmland dataset \cite{farm}, which covers a farmland area in Yancheng City, Jiangsu Province, China. The third, known as the Hermiston dataset \cite{hasanlou2018hyperspectral}, captures irrigated farmland in Hermiston City, Umatilla County, Oregon, USA.

To demonstrate the effectiveness of the proposed GLAFormer, six state-of-the-art models are selected for comparison, i.e., IR-MAD \cite{nielsen2007regularized}, SSA-SimaNet\cite{wang2021ssa}, SSCNN-S \cite{zhan2021sscnn}, CDFormer \cite{ding22grsl}, SSTFormer \cite{wang2022SSTFormer}, CSDBF \cite{wang2022csdbf} and GTMSiam\cite{wang2023gtmsiam}. GLAFormer is configured with 4 blocks and 8 attention heads across all three datasets. The comparative analysis is grounded in two primary metrics: overall accuracy (OA) and the Kappa coefficient. OA provides a general assessment of change detection performance in terms of overall correctness. Kappa is a more robust measure that takes into account the agreement between the observed classification and what would be expected by chance.

All experiments are carried out by using the Pytorch framework. The training phase spanned over 100 epochs and was conducted on a single NVIDIA 4090 GPU. The Adam optimizer is utilized with a learning rate of 0.0006. The training batch size is set as 128. The input patch size for the proposed methods is $9\times9$, and the dimension of the embedded sequence is fixed at 256. For each of the three datasets, 3\% of the samples are selected for training, 2\% for validation, and the remaining samples are used for testing.

\subsection{Experimental Results and Comparison}

\begin{table}[h]
\centering
\caption{Quantitative evaluation of different change detection methods on three datasets. The best performer is marked in \colorbox{Bg}{Blue} and the second best is underlined. The higher means better performance. }
\label{table_eva}
\resizebox{1.0\linewidth}{!}{
\begin{tabular}{ccccccc}
\hline
\textbf{Dataset}     & \multicolumn{2}{c}{River} & \multicolumn{2}{c}{Farmland} & \multicolumn{2}{c}{Hermiston} \\ \hline
\textbf{Measure}     & OA              & Kappa           & OA               & Kappa             & OA                & Kappa             \\ \hline
IR-MDA                  & 94.07           & 62.96           & 95.97            & 90.13             & 86.75             & 57.86             \\
SSA-SiamNet          & 94.87           & 72. 43          & 94.79            & 87. 43            & 93. 93            & 83.19             \\
SSCNN-S              & 95. 53          & 75.67           & 95.49            & 88. 85            & 95.69             & 87.08             \\
CDFormer             & 94.92           & 72.76           & 97.34            & 93.58             & 93.06             & 81.57             \\
SSTFormer            &96.72          &81.63          & 98.01            & 95.14             & 93.39             & 82.73             \\
CSDBF                & 95.56          & 75.61          & 98.23            & 95.75             & 96.13             & 89.51             \\
GTMSiam             &97.11          & 82.93          & 97.37            & 94.03             & 95.66             & 87.26             \\
\rowcolor{Bg} GLAFormer & \textbf{97.81}  & \textbf{83.72}  & \textbf{98.95}    & \textbf{97.14}    & \textbf{97.23}    & \textbf{91.68}    \\
 & {\color[HTML]{FF0000} ↑0.70\%} & {\color[HTML]{FF0000} ↑0.78\%} & {\color[HTML]{FF0000} ↑1.58\%} & {\color[HTML]{FF0000} ↑3.11\%} & {\color[HTML]{FF0000} ↑1.57\%} & {\color[HTML]{FF0000} ↑4.42\%}              \\ \hline
\end{tabular}}
\end{table}

As presented in Table \ref{table_eva}, the quantitative comparison between the GLAFormer and other methods is conducted on three datasets. The results demonstrate that our proposed method consistently outperforms the compared methods across all three datasets in terms of OA and Kappa. Notably, in terms of the Kappa coefficient, GLAFormer achieves an average improvement of 2.77\% across the three datasets compared to the previous state-of-the-art feature fusion method, GTMSiam. This signifies an accuracy boost of over 20\% in regions that were challenging for previous models to identify.

To demonstrate the superior performance of our proposed GLAFormer, we also qualitatively compare the results of different methods on the three datasets.  Fig. \ref{fig_river} shows the change detection result on the River dataset. Unlike other methods, the IR-MAD approach exhibits significant noise due to its lack of training information in change detection. In contrast, the detection results of GLAFormer are closer to the ground truth, particularly in the highlighted regions.
 
\begin{figure}
  \centering
  \includegraphics[width=3.5in]{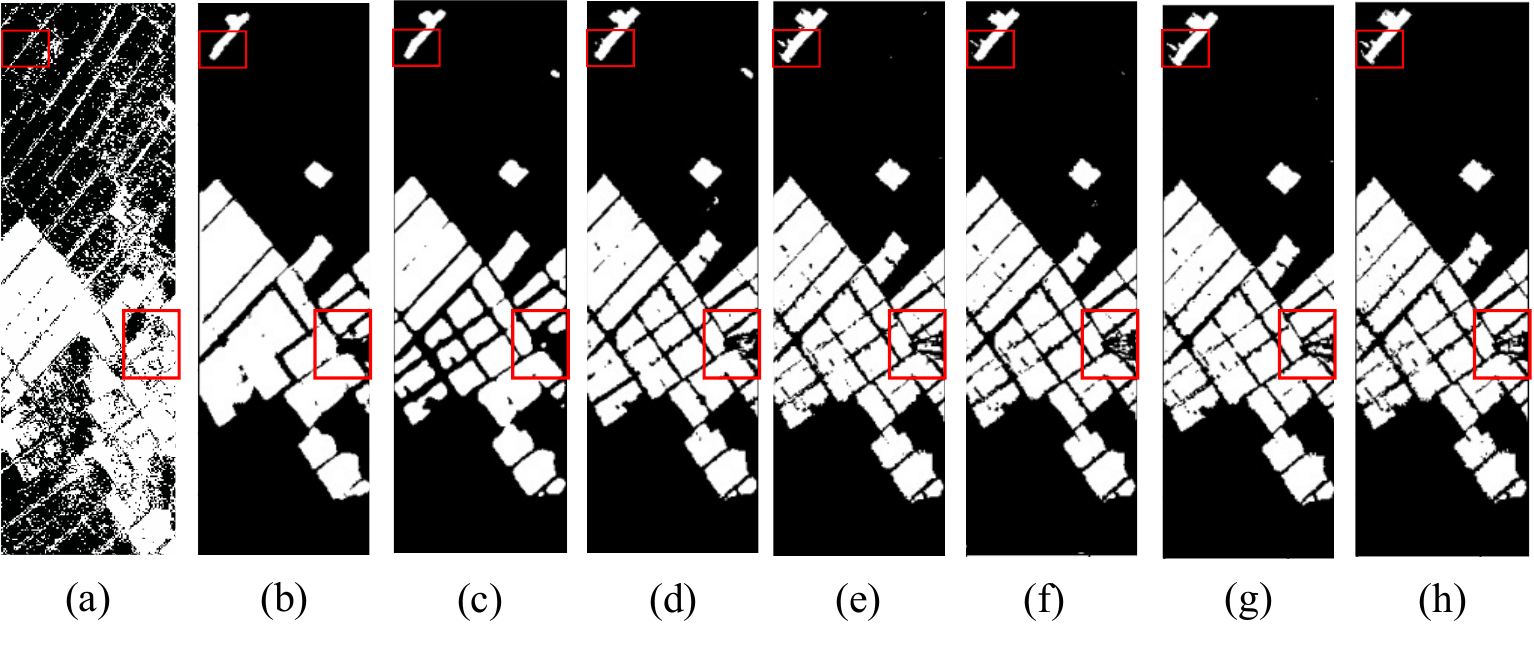}
  \caption{Change detection results of different methods on Farmland dataset. (a) IR-MAD. (b) SSA-SiamNet. (c) SSCNN-S. (d) CDFomer. (e) SSTFormer. (f) CSDBF. (g) Proposed GLAFormer. (h) Ground truth.}
  \label{fig_farm}
\end{figure}

The change detection results of the Farmland dataset is shown in Fig. \ref{fig_farm}. The regions in the red rectangle reveal that, only the GLAFormer and SSTFormer successfully identify the subtle alterations within the region. This observation aligns with the quantitative findings reported in Table \ref{table_eva}. It is worth noting that GLAFormer demonstrates an improvement of 1.58\% and 3.11\% in OA and Kappa coefficient, respectively, compared to the GTMSiam method. This performance gain can be attributed to the superior integration of global and local signals in GLAFormer.

\begin{figure}
  \centering
  \includegraphics[width=3.5in]{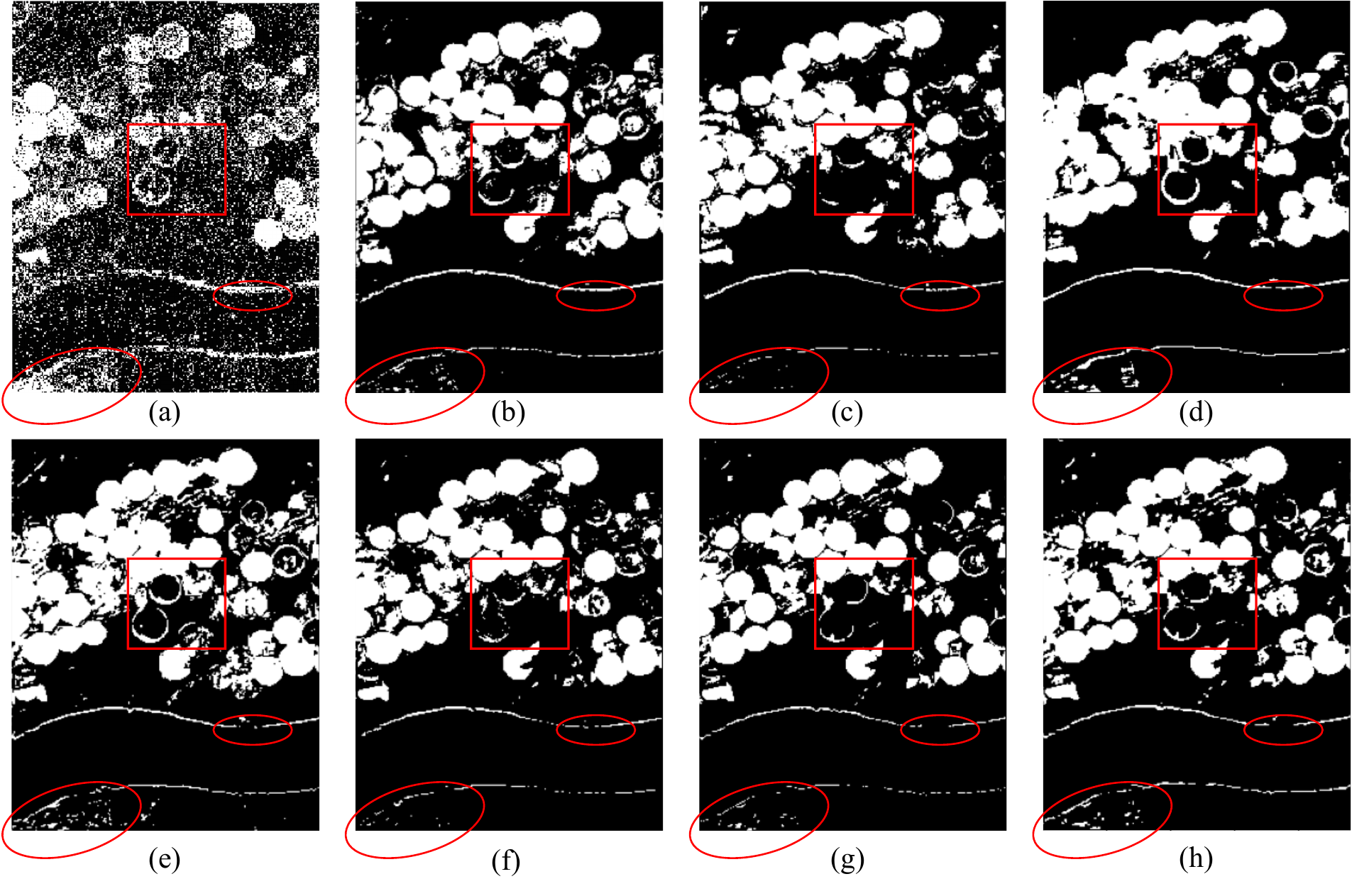}
  \caption{Change detection results of different methods on Hermiston dataset. (a) IR-MAD. (b) SSA-SiamNet. (c) SSCNN-S. (d) CDFomer. (e) SSTFormer. (f) CSDBF. (g) Proposed GLAFormer. (h) Ground truth.}
  \label{fig_sa}
\end{figure}

The change detection results of Hermiston dataset are depicted in Fig. \ref{fig_sa}. It shows the complex nature of the changes within this dataset, characterized by numerous irregular regions. A detailed analysis of the highlighted regions reveals that our proposed method produces a more accurate change map with clearer boundaries and reduced noise, underscoring its robust capability in detecting changes within complex backgrounds. This performance improvement can be attributed to the effectiveness of the dual-gated mechanism of CGFN in noise suppression. Quantitatively, GLAFormer achieves a 1.57\% and 4.42\% increase in OA and Kappa coefficient, respectively, compared to the GTMSiam method.

Both the qualitative and quantitative analysis demonstrates the robustness and accuracy of the proposed GLAFormer in handling complex change detection scenarios.

\subsection{Ablation Study}

We conduct a series of ablation experiments on the three datasets to validate the effectiveness of the proposed GLAM and CGFN. Firstly, we design a Basic Transformer with the same structure as the GLAFormer, while the Basic Transformer uses traditional multi-head attention. In addition, we have two variants of GLAFormer, i.e., without the GLAM (w/o GLAM) and without the CGFN (w/o CGFN). These are replaced with standard self-attention and FFN, respectively. The results of the ablation study are shown in Table \ref{table_ablation}. We find that GLAFormer and its variants beat the Basic Transformer in all cases. The GLAFormer always achieves better performance than its two variants on the three datasets. This demonstrates the necessity of the GLAM and CGFN designed in GLAFormer.

\begin{table}[h]
\centering
\caption{Ablation Studies of the Proposed GLAFormer. The best performer is marked in \colorbox{Bg}{Blue}.}
\label{table_ablation}
\begin{tabular}{c|c c c} 
\toprule
\multirow{2}{*}{Method}
    & \multicolumn{3}{c}{OA on different datasets ($\%$)} \\ \cmidrule{2-4}
& River & Farmland & Hermiston\\ 
\midrule
Basic Transformer    & 97.19 & 97.87  &94.42   \\
GLAFormer w/o GLAM   & 97.17  & 98.54  & 95.20    \\  
GLAFormer w/o CGFN  & 97.64  &98.66  & 96.97    \\ 
\rowcolor{Bg} Proposed GLAFormer & 97.81  & 98.95  & 97.23    \\  
\bottomrule
\end{tabular}
\end{table}

\section{Conclusions}

In this letter, we propose a novel GLAFormer for HSI change detection. The GLAFormer offers two enhancements over existing Transformer-based change detection methods. First, the designed GLAM leverages the abundant channel information intrinsic to hyperspectral images to combine both high-frequency and low-frequency signals. Furthermore, the designed CGFN is meticulously engineered to augment the extraction of pertinent information while concurrently mitigating noise interference, thereby enhancing the overall quality of the change detection process. Our comprehensive experiments conducted on three hyperspectral datasets, consistently demonstrate the superior performance of GLAFormer over the state-of-the-art methods.

\bibliographystyle{IEEEtran}
\bibliography{re} 

\end{document}